\documentclass{article}

\usepackage[english]{babel}
\usepackage{amsfonts}

\begin{document}

\title{Stability, Singularities and Mass Thresholds\\
in Child Universe Production\\
{\small{}---\hspace{2mm}\emph{a concise survey including some recent results and prospects}\hspace{2mm}---}}

\date{\hrule\vskip 2 mm%
\footnotesize{}To appear in the proceedings of\\
\emph{BH2, Dynamics and Thermodynamics of Blackholes and Naked Singularities},\\
May 10-12 2007, Milano, Italy\\
conference website: \texttt{http://www.mate.polimi.it/bh2}\\[2mm]
\hrule\vskip 2 mm%
Preprint: KUNS-2109}

\author{Stefano Ansoldi\footnotemark[1]\\
{\small{}Department of Physics, Kyoto University}\\
{\small{}and International Center for Relativistic Astrophysics (ICRA)}\\
{\small{}\textrm{email:} \texttt{ansoldi@trieste.infn.it}}\\[4mm]
Eduardo I. Guendelman\\
{\small{}Ben Gurion Univeristy, Beer Sheva, Israel}\\
{\small{}\textrm{email:} \texttt{guendel@bgu.ac.il}}\\[4mm]
Idan Shilon\\
{\small{}Ben Gurion Univeristy, Beer Sheva, Israel}\\
{\small{}\textrm{email:} \texttt{silon@bgu.ac.il}}
\footnotetext[1]{Mailiing address: Dipartimento di Matematica e Informatica,
Universit\`{a} degli Studi di Udine, via delle Scienze 206,
I-33100 Udine (UD), Italy}}

\maketitle

\begin{abstract}
We present a review of selected topics concerning the creation and evolution of child universes,
together with a concise account of some recent progress in the field.
\end{abstract}

\section{Introduction}

The questions about the origin, formation and development of our universe are
among the most challenging in physics, since they involve difficult theoretical
problems and, at the same time, are not subject to a direct \emph{experimental}
approach. Fortunately, in the last years, theoretical investigations have found
valuable allies in the increasing amount of data, which is coming from continuously
refined \emph{observational} techniques. This has given the possibility to aid
our theoretical understanding not only with a much more clear picture of the universe,
especially at the earliest stage of its evolution, but also with the results of
computational simulations, which have a consistent realization of these data as
a target of more and more refined models.

From the theoretical point of view, after the development of the inflationary scenario
\cite{bib:PhReD1981..23...347G,bib:PhLeB1982.108...389L,bib:PhReL1982..48..1220S,%
bib:MPhLA1986...1....81L,bib:PhLeB1986.175...395L,%
bib:PhReD1991..44..3152R,bib:PhLeB1994.327...208L,%
bib:PhReL1994..72..3137V},
a primary goal has been a successful consistent description of the earliest instants
of life of our universe; this problem became tightly bound with developing ideas
of vacuum and vacuum decay \cite{bib:PhReD1977..15..2929C,bib:PhReD1977..16..1248C,bib:PhReD1977..16..1762C}
about 30 years ago, when the interplay of these processes with gravitation was studied
\cite{bib:PhReD1980..21..3305L,bib:PhReD1987..36..1088W}. These studies go under
the name of \emph{vacuum bubble dynamics}: in the case of false vacuum their coupling
with gravity was firstly analyzed by Sato et \textit{al.}
\cite{bib:PrThP1981..65..1443M,%
bib:PrThP1981..66..2052M,bib:PrThP1981..66..2287S,%
bib:PhLeB1982.108....98K,bib:PhLeB1982.108...103M,%
bib:PrThP1982..68..1979S}; interesting developments
soon followed with the work of Blau \textit{et al.} \cite{bib:PhReD1987..35..1747G}
and Berezin \textit{et al.} \cite{bib:PhReD1987..36..2919T,bib:PhReD1991..43.R3112T}.
A very interesting aspect of false vacuum bubble dynamics is that it can give
rise to the formation of a child universe\footnote{The terminology \emph{child universe}
is here favored over the one \emph{baby universe} used elsewhere.}.
Child universe formation is the process in which a new universe
(the child one) emerges from an existing one (which we will call the parent universe)
in such a way that the structure of the parent universe is preserved. Since the
definition of a universe implicitly assumes that it can expand to a sufficiently large
size and live for a sufficiently long time (so that in it structures similar to the one
that we observe in our universe can be formed) child universe formation (as defined above)
may seem impossible. This is, instead, not the case in a general relativistic framework,
which allows a rich enough structure in the causal structure of spacetime to satisfy
both the above conditions. Diverse realizations of the child universe creation process
are possible and we will provide a short review of some of them below: for concreteness,
our attention will concentrate on the model originally developed in
\cite{bib:PhReD1987..35..1747G} as well as later refinements of it.
In this context, the importance of \cite{bib:PhReD1987..35..1747G} resides in the fact that,
for the first time, geodesically complete coordinate systems were used to describe the
universe formation process. This, apparently purely technical, point has instead direct
advantages in the physical interpretation of the model, since it emphasizes the crucial
role played by wormholes. As we will also repeat below, if we consider a small bubble of false
vacuum which expands to a very large size (the baby universe) the energy density inside the
bubble is higher than the energy density outside it; the mechanical force is then directed
from the parent space toward the baby universe. On one side of the wormhole this is the
direction of decreasing radius (so that in this case there can be no expansion). But if the
solution can evolve on the other side of the wormhole, the opposite situation
is realized, i.e. the mechanical force pushes in the direction of increasing radius. Thus, the
false vacuum bubble can (and will) expand; from the point of view of the parent spacetime,
this process is taking place on the other side of the wormhole throat. Thus the growth
of the child universe takes place without affecting in any way the evolution of the
parent universe, which witnesses at most the formation of a black hole\footnote{These ideas
can be generalized to situations in which the black hole is created as a result of gravitational
collapse (see, for instance, \cite{bib:PhReD1996..53..3215F,bib:PhLeB2001.506...351G} and references
therein; more references can also be found in the additional contribution of S. Ansoldi
to the proceedings of this conference).}.
The child universe, in fact, grows \emph{creating its own space}.

After introducing in a very concise way some necessary background formalism, with the intent
of fixing notations and conventions (section \ref{sec:backgr}), in section \ref{sec:vacdecbubchiuni}
we present a (non exhaustive) review of some of the above ideas. In particular, in three separate
subsections we revisit some features of child universe formation: in
subsection \ref{subsec:sinavoandtun} we analyze the occurrence of singularities
and the problems with the tunnelling process that can be used for singularity avoidance
in semiclassical, minisuperspace models; in subsection \ref{subsec:sta} we review the stabilization
of the initial (i.e. \emph{pre}-tunnelling) configuration, both from the classical and semiclassical
standpoint; in subsection \ref{subsec:crimas} we discuss the presence of a critical mass threshold for
child universe formation as well as early proposals to reduce it or trade it with other properties
of spacetime and/or its matter content. Section \ref{sec:recmodposdev} follows, where we describe
a recent realization of child universe production free of a mass threshold; in this model child universe
production can, surprisingly enough, take place \emph{out of almost empty space}, a fact that is
quite suggestive, especially in the perspective of unsuppressed transplanckian child universe production,
which has also been recently discussed. We conclude the paper in section \ref{sec:conclu}, with a
synopsis and a concise discussion/remark, stressing again what is the main idea that we have revisited
in the rest of the paper.

\section{\label{sec:backgr}Background formalism: an essential review}

We will here recall more details related to the general discussion that we made in the introduction
and review various interesting developments that appeared in the literature following the ideas
proposed in \cite{bib:PhReD1987..35..1747G}. In particular we will consider the description of
vacuum bubble in terms of general relativistic shells, restricting ourself to the spherically symmetric
case. The dynamics of a generic shell of matter-energy, $\Sigma$, can be described by Israel junction
conditions \cite{bib:NuCim1966.B44.....1I,bib:NuCim1967.B48...463I,bib:PhReD1991..43..1129I}. The shell
is the common part of the boundaries of the two spacetime manifolds ${\mathcal{M}} _{\pm}$ that it joins,
i.e. $\partial {\mathcal{M}} _{-} \cap \partial {\mathcal{M}} _{+} = \Sigma$. Geometrically the embedding
of $\Sigma$ in ${\mathcal{M}} _{\pm}$ is described by the corresponding extrinsic curvatures\footnote{In
the following Greek indices $\alpha$, $\beta$, $\mu$, $\nu$, $\dots$ take the values $0$, $1$, $2$, $3$ and
Latin indices $a$, $b$, $i$, $j$, $\dots$ take the values $0$, $2$, $3$.}
$K ^{(\pm)} _{ij}$ and, for a non-lightlike junction, the junction conditions can be expressed
in terms of the jump of the extrinsic curvature across the shell
\[
    [ K _{ij} ] \stackrel{\mathrm{def.}}{=} K ^{(+)} _{ij} - K ^{(-)} _{ij}
    ,
\]
which is related to the shell stress energy tensor $S _{ij}$ by \cite{bib:NuCim1966.B44.....1I,bib:NuCim1967.B48...463I}
\begin{equation}
    [ K _{ij} ] = 8 \pi G S _{ij}
    .
\label{eq:isrjuncon}
\end{equation}
Moreover the conservation equations imply
\[
    S ^{i} _{j;i} = [ e ^{\alpha} _{(j)} T ^{\beta} _{\alpha} n _{\beta}]
\]
where $T ^{(\pm)} _{\mu \nu}$ describes the energy-matter content of ${\mathcal{M}} _{\pm}$,
$n ^{\mu}$ is the normal to the shell $\Sigma$, which we assume pointing from ${\mathcal{M}} _{-}$ to ${\mathcal{M}} _{+}$,
and $e ^{(\pm) \alpha} _{(i)}$ are the components of a basis in the tangent space to $\Sigma$ when evaluated
in ${\mathcal{M}} _{\pm}$, respectively.
When spherical symmetry is enforced the system of equations (\ref{eq:isrjuncon}) contains only one independent
equation. For definiteness let us choose in the two spacetimes the metric tensor in the form which is static
and adapted to the spherical symmetry, so that
\[
    g ^{(\pm)} _{ij}
    =
    \mathrm{diag} ( - f _{(\pm)} (r _{(\pm)}) , f ^{-1} _{(\pm)} (r _{(\pm)}) , r _{(\pm)} ^{2} , r _{(\pm)} ^{2} \sin ^{2} \theta _{(\pm)} )
\]
if we choose the coordinates $(t _{(\pm)} , r _{(\pm)} , \theta _{(\pm)} , \phi _{(\pm)})$. It is not restrictive, in view of the
spherical symmetry, to assume $\theta _{(-)} = \theta _{(+)} = \theta$ as well as $\phi _{(-)} = \phi _{(+)} = \phi$. The shell will
then be described by its radius $R ( \tau )$, described as a function of the proper time $\tau$ of an observer comoving
with the shell. We then have that $R (\tau)$ is the only independent function describing the intrinsic metric on the shell.
In view of
\[
     R ( \tau) = r _{(\pm)} \rceil _{\Sigma}
    ,
\]
the continuity of the three metric across $\Sigma$ is then realized, as requested by Israel junction conditions. Within the above
settings the only remaining junction condition, mentioned above, can be written as
\begin{equation}
    \epsilon _{(-)} \sqrt{\dot{R} ^{2} + f _{(-)} (R)}
    -
    \epsilon _{(+)} \sqrt{\dot{R} ^{2} + f _{(+)} (R)}
    =
    \frac{G M(R)}{R}
    ,
\label{eq:sphjuncon}
\end{equation}
where an overdot represents a derivative with respect to $\tau$ and $M (R)$ describes the shell matter content after
a suitable equation of state has been fixed. The quantities $\epsilon _{\pm}$ are signs, related to the direction
of the normal in the maximal extension of the spacetimes described by the metrics $g ^{(\pm)} _{\mu \nu}$.

\section{\label{sec:vacdecbubchiuni}Vacuum decay, bubbles and child universes}

The above summarized formalism has been widely used to describe the formation of baby/child universes and
is a convenient practical way to implement the more detailed description in which the vacuum decay
is, in fact, represented by the decay of a scalar field. If the classical field has two equilibrium
states which are both classically stable but have a different energy density, when quantum effects are
switched on the higher energy density state can become unstable because of the possibility of tunnelling
under the potential barrier. The interesting effects that may arise are already transparent in
the semiclassical approximation when a single scalar field with nonderivative interactions is considered:
the decay of a volume ${\mathcal{V}}$ of the higher energy density false vacuum into the lower energy density
true vacuum has a probability (per unit time per unit volume) given by
\[
    \frac{\Gamma}{{\mathcal{V}}} = A e ^{B / \hbar} \left[ 1 + {\mathcal{O}} (\hbar) \right]
\]
and the exponent $B$ \cite{bib:PhReD1977..15..2929C} as well as the coefficient $A$
\cite{bib:PhReD1977..16..1762C} can be computed with (now) standard techniques.

The above analysis acquires particular relevance when it comes into contact with gravitation. Indeed
in Einstein's theory we have a strong connection between properties of space time and matter, and the
concept of vacuum acquires a more subtle meaning. It was early recognized \cite{bib:PhReD1980..21..3305L}
that a study of vacuum that would not include gravitational effects would have been seriously incomplete
and the effect of gravitation upon the decay of false vacuum
was, then, studied \cite{bib:PhReD1980..21..3305L}. Particular attention was given to the cases
in which the so called ``thin-wall approximation'' was satisfied, a situation in which
the thin-shell formalism briefly summarized in the previous section can also be used
to describe this process.

The vacuum bubbles described above are true vacuum bubbles, which arise in a midst of false vacuum.
It was early recognized that the opposite process is also possible, i.e. false vacuum bubbles can
be created in a midst of true vacuum. What makes these bubbles interesting in the cosmological
context, is that they undergo an exponential growth, a fact that recalls the exponential inflation
of the universe during the inflationary era. The classical behavior of false vacuum bubbles
was also early analyzed \cite{bib:PrThP1981..65..1443M}. In the original model the universe was
modelled by an inner false vacuum core, surrounded by a shell-like true vacuum region, which was itself
immersed in a false vacuum ``bath''. In this model the vacuum phase transition appears with an interesting
connection to the formation of black holes/wormholes \cite{bib:PhReD1987..35..1747G,bib:MPhLA1991...6..1535T}
in the spacetime structure (a fundamental point, as we will
see later on). A detailed analysis of the false vacuum bubble creation process soon appeared
\cite{bib:PrThP1981..66..2052M}, followed by a generalization in which black holes are also magnetized,
a relevant aspect in first order GUT phase transitions \cite{bib:PrThP1981..66..2287S}. A natural
model of a false vacuum bubble including gravitational effects is obtained choosing de Sitter spacetime
to describe the inside of the bubble and Schwarzschild spacetime to describe the outside.
Generalizations, of course, are possible, and, for instance, Schwarzschild--de Sitter spacetime
can be considered to extend the model \cite{bib:PhLeB1982.108....98K}: the possibility of
multi-production of child universes (an early formulation of a very actual idea in cosmology), i.e.
the fact that the production of child universes will continue indefinitely giving rise to
a hierarchical structure of \emph{universes inside universes}, also appeared \cite{bib:PhLeB1982.108...103M}.
The relevance of thermal bubble nucleation, in conjunction with the quantum production process, has also been
emphasized, together with the important role that is played by primordial black holes and wormholes
\cite{bib:PrThP1982..68..1979S}, a point which will be central in our following discussion\footnote{Incidentally,
we would like to remark, in a historical perspective, that the general picture of the universe implied
by these early models \cite{bib:PrThP1982..68..1979S} is very close to the currently accepted point of view,
which is nowadays suggested by alternative motivations, like those coming from string theory.}.

The most common approach to the formation of inflating bubbles that we have described up to now
has been mostly field-theoretical in nature (see also the analysis in \cite{bib:PhReD1987..36..1088W}).
There is, nevertheless, a way to analyze the same problem at the classical level focusing directly on
general relativistic oriented description \cite{bib:PhReD1987..35..1747G,bib:PhReD1987..36..2919T}:
the quantum generalization of these ideas would provide an interesting refinement of the analysis
already developed in \cite{bib:PhReD1984..30...509V}, where in the minisuperspace approximation the
wave function describing quantum creation of the universe from nothing is studied. Following for definiteness
the analysis developed in \cite{bib:PhReD1987..35..1747G}, it is possible to see that the formalism described
in section \ref{sec:backgr} is perfectly suited to analyze the formation of a vacuum bubble. Moreover,
an immediate interpretation of the properties of the model is made possible by the intuitive geometrical
meaning of the quantities which appear in the junction condition (\ref{eq:sphjuncon}). In more detail
the model studied in \cite{bib:PhReD1987..35..1747G} describes the dynamics of a vacuum bubble in
terms of the dynamics of a general relativistic shell separating a region of de Sitter spacetime,
\[
    \mathrm{i.e.}
    \quad
    f _{(-)} = 1 - \chi ^{2} r _{(-)} ^{2}
    \quad
    \mbox{in the notation introduced in section {\protect\ref{sec:backgr}}}
    ,
\]
from a region of Schwarzschild spacetime,
\[
    \mathrm{i.e.}
    \quad
    f _{(+)} = 1 - 2 G m / r _{(+)}
    \quad
    \mbox{in the notation introduced in section {\protect\ref{sec:backgr}}}
    .
\]
The energy-matter content of the shell corresponds to the presence of a uniform, positive,
energy-density $\sigma$, which equals the opposite of the (equal) radial and tangential
pressures. Thus, if $h _{ij}$ is the induced metric on the shell, we have
\[
    S _{ij} \propto \sigma h _{ij}
    \quad
    \mbox{and correspondingly}
    \quad
    M (R) \propto 4 \pi \sigma R ^{2}
    .
\]
This equation of state naturally arises from the above described field theoretical models
using a scalar field to describe the vacuum phase transition. The study of the solutions
of the junction condition (\ref{eq:sphjuncon}) can then be performed in this set-up. The
apparent difficulties due to the unusual structure of (\ref{eq:sphjuncon}) are, in fact,
easily dealt with resorting to an equivalent effective formulation that reduces the problem
to the analysis of the motion of a classical particle of unit mass in a given
potential\footnote{See, for instance, \cite{bib:PhReD1987..35..1747G} as well as
\cite{bib:PhReD1989..40..2511S}, where the more general Schwarzschild--de Sitter
exterior geometry is used. A recent survey of the properties of the effective
description for general situations, where specific forms of $f _{(\pm)} (r _{\pm})$
and of the matter content $M (R)$ are not assumed, can be found in
\cite{bib:QGF072007...1...004A}.}. The results about the behavior of the radial
coordinate $R (\tau)$ have to be complemented with the determination of the
$\epsilon _{(\pm)}$ signs, which can also be obtained in closed form in the
general case \cite{bib:QGF072007...1...004A}.

The parameters of this model are three\label{top:orimodthrpar}, i.e.
the cosmological constant of the false vacuum bubble $\Lambda = 3 \chi ^{2}$
(which can be connected to the vacuum energy density $\varepsilon$ by the relation
$\Lambda = 8 \pi G \varepsilon$),
the Schwarzschild mass of the asymptotically flat region, $m$, and the surface energy
density $\sigma$. It can be seen that, varying these parameters, only two
qualitatively different situations for the solutions of (\ref{eq:sphjuncon})
are obtained:
\begin{enumerate}
    \item the system admits solutions which starts from zero radius $R$ and grow up
    to arbitrary large $R$; these solutions grow enough to describe the formation and
    subsequent evolution of an inflationary universe, but develop from an initial
    singularity;
    \item the system admits two kinds of solutions:
    \begin{enumerate}
        \item the so called \emph{bounded solutions}, which start evolving from zero radius and, after reaching
        a maximum radius of expansion, collapse back to zero radius;
        \item the so called \emph{bounce solutions}, which collapse from large values of the radius up to a
        minimum radius and then expand again to infinity;
    \end{enumerate}
    in this situation, in which the classical transition from a bounded to a bounce solution is forbidden
    by the existence of a potential barrier, the past-singularity--free bounded solutions cannot become big
    enough to describe the evolution of a universe like ours; at the same time, the bounce solutions, cannot
    evolve in the small radius region, i.e. they are not appropriate to describe the early evolution of a universe
    like ours.
\end{enumerate}
As we will discuss in more detail below, it seems thus that the only choice to have a bubble evolution that is suitable
for the description of the observed universe is to accept the initial singularity.
This result is not a failure of the present model, but is connected with general results
of singularity theorems in general relativity \cite{bib:PhReL1965..15...689H,bib:PRSLA1966.294...511H,%
bib:PRSLA1966.295...490H,bib:PRSLA1967.300...187H,bib:PRSLA1970.314...529P}
(see also \cite{bib:CaUPr1973...1...391E} and \cite{bib:UofCP1984...1...491W} for standard textbooks
on the subject) as it was early recognized \cite{bib:PhLeB1987.183...149G}. Granted this,
\emph{classical models} are still very useful to investigate the essential features of
the child universe creation process, as we will now briefly discuss.

In particular, we remember that the child universe is characterized by an energy density which is
higher than that of the surrounding parent universe. Thus it has lower pressure and it is, then,
impossible for it to exert an expanding action in the direction of increasing radius, i.e.
to expand filling the parent universe. There is, nevertheless, another way in which the child universe
\emph{can} expand, which relies on the peculiar properties of the parent spacetime according to
general relativity: if a wormhole is present, then the child universe, can grow by
\emph{making its own space} on the other side of the wormhole throat. In this case the late time
evolution of the child universe is \emph{classically} hidden from an observer in the parent universe
by the presence of an event horizon, i.e. after the initial stage of the evolution, the child universe
disconnects from the parent one and, independently, continues its expansion, growing into a full
universe. Observers in the parent universe will, thus, only witness its birth (at least classically)
by looking very far into their past. It is worth to remark that it is exactly the wormhole structure
of spacetime that allows for the creation of the child universe: the same force that would prevent the
expansion on the side of the wormhole where observers living in the parent universe evolve, acts
in the opposite direction, i.e. favors the expansion, on the other side of the wormhole. In this
region the normal, which is directed from the newborn child universe toward the parent space, points in the
direction of decreasing radius: mathematically this is reflected by $\epsilon _{(+)} = -1$ in equation
(\ref{eq:sphjuncon}). If this condition is met for values of the parameters for which solutions of type
1 above can be realized, i.e. for which classical solutions starting from zero radius and expanding
to infinity do exist, then child universes are formed\footnote{We will discuss later on the issues related
to the purely classical creation process, as well as the ideas that have been developed to circumvent them.}.
Summarizing, two requirements have to be met for child universes to be realized as false vacuum bubbles
using the thin-wall approximation:
\begin{description}
    \item[requirement 1:\label{top:chiunicrereq001}]
    there must be a process by which a \emph{very small} bubble can become
    \emph{big enough}, so that, both, the early and late time evolution of our universe can be described;
    \item[requirement 2:\label{top:chiunicrereq002}]
    at late time, the evolution must guarantee that $\epsilon _{(+)} = -1$, so that
    the universe is, effectively, classically disconnected from the parent universe and expands by creating
    \emph{its own space}.
\end{description}
The above requirements can be applied also to generalizations of the model, which was firstly analyzed in
\cite{bib:PhReD1987..35..1747G}, and also to those generalizations involving semiclassical quantum effects.

The original model that we have reviewed above is very immediate in its description of the child universe
formation process, but is not completely satisfactory. There are, in fact some features that deserve a closer
investigation and that might suggest refinements and/or generalizations. In what follow we will be interested
specifically in the following ones:
\begin{enumerate}
    \item the presence of singularities: according to general theorems, as mentioned above, the classical models
    existing in the literature are affected by the presence of an initial singularity, which is unlikely to
    disappear within the framework defined by the classical treatment;
    \item the problem of stability: the baby universe configurations might be unstable and a similar problem could
    appear when, in trying to address the singularity problem, quantum effects are invoked; in this second case, the
    stability of the initial configuration has also to be analyzed;
    \item the presence of a critical mass: in most of the models present in the literature the two requirements
    listed above as necessary for the formation of a child universe, are usually both satisfied at the classical
    level only if a threshold on the total mass energy of the parent spacetime is satisfied.
\end{enumerate}

\subsection{\label{subsec:sinavoandtun}Singularity avoidance and tunnelling}

Let us start with the first of the issues mentioned above, i.e. the one of singularities. As we already said, this
is a general problem, related to singularity theorems of general relativity, which is very difficult to
avoid at the purely classical level \cite{bib:PhReL1965..15...689H,bib:PRSLA1966.294...511H,%
bib:PRSLA1966.295...490H,bib:PRSLA1967.300...187H,bib:PRSLA1970.314...529P,%
bib:CaUPr1973...1...391E,bib:UofCP1984...1...491W}. Recently there has been a proposal of a classical model,
in which the spacetime is singular, but the singularity does not appear in the causal past of the newly
forme universe \cite{bib:PhReD2006..74024026K}, which could be called a \emph{weak} singularity avoidance.
The model is based on a magnetic monopole, a theme that will appear again later on. Anyway, if we insist
to have a completely regular spacetime, in which all energy conditions are satisfied, this can not be achieved
at the purely classical level\footnote{It is possible, of course, to relax some of the energy conditions,
for instance the strong one. This is the subject of some interesting ideas, which have been developed
in connection with regular black holes spacetime. See the additional contribution of S. Ansoldi to the
proceedings of this conference for a concise review of \emph{black hole spacetimes with a regular center}
and additional references.}.
This is why, very early, quantum effects have been proposed as a possible way to solve this problem. This is the early
proposal that can be found in \cite{bib:NuPhy1990B339...417G} and in
\cite{bib:PhReD1990..41..2638P,bib:PhReD1990..42..4042P}. This proposal moves from the observation,
recalled above, that in the presence of potential barrier neither the bounded nor the bounce solutions
are appropriate models to describe the formation and subsequent evolution of an inflationary universe
like our one. On the other hand, if tunnelling could be allowed, then, starting from the nonsingular
bounded solution, the universe could tunnel quantum mechanically under the potential barrier and
continue its evolution along the expanding branch of the bounce trajectory. Singularities would then
be avoided and it can be shown that a baby universe can be formed in the process. Despite the early
appearance of this model, it is remarkable that some of the difficulties, which emerged in the initial
formulation \cite{bib:NuPhy1990B339...417G}, are still not yet solved \cite{bib:FFP6p2006........69S,%
bib:PhReD2005..72103525J,bib:PhReD2006..73123529J,bib:JGR162007...1...114A,bib:QGF072007...1...004A}.
In particular it was shown in \cite{bib:NuPhy1990B339...417G} that it is not always possible to construct
in a consistent way the Euclidean manifold which interpolates between the bounded and bounce classical
configurations. Moreover, in these situations, path-integral and canonical quantization give different
results: although it may be argued that the path-integral method is the more consistent, it would be,
nevertheless, interesting to understand the precise reasons for the failure of the canonical one.
Last but not least, we would like to mention that:
\begin{itemize}
\item[--] although a Lagrangian formalism exists (see for instance \cite{bib:PhReD2005..71064008I}
and references therein), which can be reduced to an effective spherically symmetric Lagrangian formalism
\cite{bib:ClQuG1997..14..2727S},
\item[--] despite the fact that this formalism can be effectively used to reproduce \cite{bib:ClQuG1997..14..2727S}
the results in \cite{bib:PhReD1980..21..3305L} and their later generalizations \cite{bib:PhLeB1983.121...313P},
\end{itemize}
\noindent{}when some problems appear in the structure of the Euclidean manifold, as discussed above, the Euclidean
momentum, which is obtained from the effective Lagrangian, cannot be defined in such a way that it is, both,
vanishing at the turning points and continuous along the tunnelling trajectory \cite{bib:FFP6p2006........69S,%
bib:JGR162007...1...114A,bib:QGF072007...1...004A}. These facts signal that something is still
not completely understood in the tunnelling process.

\subsection{\label{subsec:sta}Stability}

Connected with the issue above is the problem of stability, since, following the proposal of implementing
quantum effects (at least at the semiclassical level) to describe the singularity-free birth of a child-universe
\cite{bib:NuPhy1990B339...417G,bib:PhReD1990..41..2638P,bib:PhReD1990..42..4042P}, it turned out that the
initial configuration would have been unstable. A possible way to address this issue, which is interesting
also from other perspectives, is by the addition of a more elaborate matter content on the bubble surface,
in addition to the standard tension term, which is suggested by field theoretical approaches. In particular,
it was shown in \cite{bib:ClQuG1999..16..3315P} that the addition of gauge fields gives the possibility
to obtain a classically stable initial configuration, which may be metastable, from which the quantum
tunnelling process might, then, initiate. In the case which we discussed above, in which a de Sitter--Schwarzschild
junction is performed by a shell having a uniform tension that equals the opposite of its energy density, the
generic form of the potential diverges to minus infinity at both small and large value of $r$. The addition
of a gauge field content on the bubble surface, modifies the small scale behavior, adding a potential well
corresponding to a stationary classical configuration. An analogous result can be obtained invoking
quantum effects for the stabilization process: in particular the bounded solution can be also considered
in a quantum regime, and quantized semiclassically, using for instance the Bohr-Sommerfeld quantization
condition \cite{bib:ClQuG2002..19..6321A,bib:AIPcp2005.751...159A}. Although the junction performed in
\cite{bib:ClQuG2002..19..6321A} is not the most relevant in connection with cosmological applications,
to reiterate the analysis in phenomenologically more sound situations does not pose any new technical problems.
In this way it could then be possible to obtain a whole set of states, the ground state as well as excited
ones, that might represent a suitable, metastable, quantum configuration to be used as initial configuration.
It has, moreover, been shown that, at least in lower dimensional models, the amount of tunnelling required
can be arbitrarily small \cite{bib:MPhLA2001..16..1079P}.

\subsection{\label{subsec:crimas}Critical mass}

As a last point, we discuss the critical mass feature, which characterizes many models. In particular,
it can be seen that for the universe creation process to take place, the mass parameter of the parent
spacetime, which is modelled by the Schwarzschild solution, must be greater than a critical value,
$m _{\mathrm{cr}}$. The possibility to lower arbitrarily the mass threshold, has been considered
and it has been proved that it can be realized, provided some scalars fields are added to
the model \cite{bib:PhReD1991..44..3152R}. The scalar fields must have a ``hedgehog'' configuration,
more precisely the configuration of a global magnetic monopole of big enough strength\footnote{For other works discussing
the subject of topological inflation see \cite{bib:PhReL1994..72..3137V,bib:PhLeB1994.327...208L},
where the gauged case is analyzed in detail.}.
Important consequences then appear when considering, both, true as well as false vacuum bubbles. When true
vacuum bubbles are considered, the addition of an ``hedgehog'' is relevant in connection with vacuum stability,
otherwise, i.e. in the case of false vacuum bubbles, for cosmological applications:
we are interested in this second possibility. From the point of view of the gravitational properties of the
system\footnote{The gravitational effect of a global hedgehog in Einstein's theory has been
originally studied in \cite{bib:PhReL1989..63...341V}.}, the addition of a ``hedgehog'' in the model described
in the previous section corresponds to the following choice \cite{bib:PhReD1991..44..3152R}
for the function $f _{(-)} (r _{(-)})$, which appears in equation (\ref{eq:sphjuncon}):
\[
    f _{(-)} (r _{(-)}) = \kappa - \chi ^{2} r ^{2} _{(-)}
    \quad \mathrm{where} \quad
    \kappa = 1 - 8 \pi G \mu ^{2}
    ,
\]
$\mu$ being the strength of the magnetic monopole.
Then, a big enough value of $\mu$ assures that all bubbles starting from zero radius can expand up to infinity,
since the role of the spatial radial coordinate and of the time coordinate is interchanged: then the radial
coordinate becomes timelike and can only increase, driving the expansion.
The possibility to dynamically obtain a configuration with big enough $\mu$ has also been discussed
\cite{bib:PhReD1999..59043513V}. In particular monopole collision has been analyzed as an example
of a process that can create a supercritical monopole and is analogous to gravitational collapse;
a slightly different scenario, involving the topological inflation of a magnetic monopole in the early
universe has been analyzed as well. Further developments of these and of earlier ideas, where
magnetic monopoles are also relevant in a cosmological context
\cite{bib:PhReD1996..53...655M,bib:PhReD1996..54..1548S,bib:PhReD1997..56..7621V},
have recently appeared \cite{bib:PhReD2006..74024026K}, in connection with, both, a purely
classical as well as a semiclassical analysis.

With the above discussion, we have tried to substantiate both the importance as well as the difficulties
in the description of child universe formation. Some of the issues, which are quite transparent already
in the simplest models in $4$-dimensions, are, in our opinion, a clear signal that we still miss some
essential features of the child universe formation process. Moreover, it seems also clear that the solution of
these issues is instrumental for a deeper understanding of the properties of the early universe and
can be a useful laboratory to test our understanding of the interplay between gravitation and the quantum
realm. This said, in what follows, we are going to briefly outline some recent results, which suggest
a relevance of the child universe formation process even beyond the domain of early universe cosmology,
and show that the transplanckian creation of baby universes might be unsuppressed and take place out
of almost empty space.

\section{\label{sec:recmodposdev}A recent model and its possible developments}

A technically simple model for baby universe creation can be, in fact, obtained
even in a framework which is more simplified than the one of the models discussed above.
Let us choose, in particular, a model characterized by the ${\mathcal{M}} _{-}$
metric with
\[
    f _{-} (r _{(-)}) \equiv 1
    , \quad
    \mbox{i.e. Minkowski spacetime}
\]
and by the ${\mathcal{M}} _{+}$ metric
\[
    f _{+} (r _{(+)}) \equiv 1 - \frac{2 G m}{r _{(+)}}
    , \quad
    \mbox{i.e., again, Schwarzschild spacetime}
    .
\]
From the very interesting ideas proposed in \cite{bib:ClQuG1999..16..3315P}, and briefly mentioned
above, it can be seen that an important role in shell based models is played not only by the
spacetime structure of the manifolds ${\mathcal{M}} _{\pm}$, but also by the matter content of the
shell. In \cite{bib:gr-qc2007..06..1233G} it has been proposed to consider the equation of state
\[
    p = - \frac{\rho}{2}
    ,
\]
$p$ being a uniform isotropic pressure on the shell and $\rho$ being its uniform energy density.
Since the shell is, spatially, a two dimensional object (a sphere, ${\mathbb{S}} ^{2}$) and in
view of the fact that a string gas \cite{bib:NuPhy1989B316...391V,bib:AstrJ1989.344...543K,%
bib:PhReD2000..62103509E,bib:NuPhy2002B623...421K} in $n$ spatial dimensions obeys the equation of state
$p = - \rho / n$, we can interpret the matter content of the shell in this model as a gas
of strings confined on the shell\footnote{Note that a connection between
global strings, shells and formation of false vacuum bubbles was already
drawn in \cite{bib:PhReD1991..44..3152R}.}.
Correspondingly, we have
\[
    M (R) = c R
    , \quad \mathrm{where} \quad
    c = 4 \pi \rho _{0}
\]
and $\rho _{0}$ is related to the energy density $\rho$ by $\rho = \rho _{0} / R$. As discussed in
\cite{bib:gr-qc2007..06..1233G} this model realizes child universe creation if
\begin{equation}
    c > \frac{2}{G}
    \quad \mathrm{i.e.} \quad
    \rho _{0} > \frac{1}{2 \pi G}
\label{eq:chiunicrecon}
\end{equation}
i.e. if there is enough amount of strings on the shell. This can be seen easily by studying the
value of the sign $\epsilon _{(+)} = \epsilon _{(+)} (R)$ for classical solutions as the parameters
$m$ and $c$ are varied. It turns out that at small $R$ the behavior is always dominated by a term
coming from the first inverse power behavior of $- f _{(+)}$, whereas at large $R$ the dominant contribution
is coming from the opposite of the square of the linearly diverging contribution coming from $M (R)$.
Thus at small $R$ we have $\epsilon _{(+)} (R) = +1$ and at big $R$, instead, $\epsilon _{(+)} (R) = -1$.
This last results is in agreement with the second requirement for universe creation that we discussed
on page \pageref{top:chiunicrereq002}. It is also straightforward to study the effective potential
\cite{bib:PhReD1987..35..1747G,bib:PhReD1989..40..2511S,bib:ClQuG1997..14..2727S} of this model
\cite{bib:gr-qc2007..06..1233G}, and see that it
is a monotonically increasing function of $R$, tending to minus infinity when $R$ tends to
zero and tending to the value $1 - G ^{2} c ^{2} / 4$ from below when $R$ tends to plus infinity.
Then, as we anticipated above, also the first requirement for child universe creation that
we discussed on page \pageref{top:chiunicrereq002} can be satisfied, provided $c > 2 / G$.
We remember that the total mas-energy of the shell is given by $M (R)$, i.e. it is
\[
    M ( R ) = c R = 4 \pi \rho _{0} R =  \frac{\rho _{0}}{R} (4 \pi R ^{2})
    ,
\]
so that its surface energy density is given by
\[
    \rho = \frac{\rho _{0}}{R}
    .
\]
On the other hand
\[
    M ( R ) = (\mbox{number of strings}) \times (2 \pi R) \times (\mbox{string tension})
\]
so that, by direct comparison, the parameter $\rho _{0}$ can be interpreted as
\[
    \rho _{0} = \frac{1}{2} \times (\mbox{number of strings}) \times (\mbox{string tension})
\]
and we see that condition (\ref{eq:chiunicrecon}), which guarantees child universe production,
requires the total string content of the shell to be transplanckian.

In the above discussion the relevant results do not depend on the total mass parameter of
the parent spacetime $m$; moreover, we remember that the new created universe is a flat,
open, Minkowski universe. We thus see that the model allows child universe creation out
of almost empty space. This result can have interesting implications, as discussed in
\cite{bib:gr-qc2007..06..1233G}, and is subject to various possible refinements,
which will have similar properties \cite{bib:gr-qc2007.105....03G}.

Of course, the model described above is rather simplified: generalizations are desirable,
perhaps required, and are under consideration. We will, thus, conclude this section
by reporting briefly about this work in progress. A guiding line in seeking more general
settings that allow child universe production out of almost empty space can be the analysis
presented in \cite{bib:gr-qc2007.105....03G}, where it is shown that the mass threshold
\cite{bib:PhReD1987..35..1747G,bib:NuPhy1990B339...417G}, $m _{\mathrm{cr}}$, for child universe
production can be arbitrary small. As we reviewed in the central part of section \ref{sec:vacdecbubchiuni},
on page \pageref{top:orimodthrpar},
the model presented in \cite{bib:PhReD1987..35..1747G} has three parameters, the false vacuum
bubble energy density, $\varepsilon$ (that can be modelled by a cosmological constant $\Lambda$ in ${\mathcal{M}} _{-}$),
the string/shell energy density/tension, $\sigma$, and the parent spacetime total
mass, $m$. In \cite{bib:gr-qc2007.105....03G} it is shown that when $\varepsilon$ and/or $\sigma$ tend
to higher and higher values, then the threshold on $m$ for child universe production, $m _{\mathrm{cr}}$,
becomes smaller and smaller, so that a lower and lower amount of quantum tunnelling \cite{bib:NuPhy1990B339...417G}
is required: thus child universe production probability is enhanced. Clearly, it is important to generalize
the classical, two parameter model that is described in \cite{bib:gr-qc2007..06..1233G}, to more general
situations. A first step is to allow for a de Sitter space, instead than the Minkowski one, for the
${\mathcal{M}} _{-}$ spacetime, i.e. to add a vacuum energy density to the model. In this case, it
can be proved that condition (\ref{eq:chiunicrecon}) for child universe creation \emph{does}
in fact persists \cite{bib:IPREP2007ShiGueAns..}. Generalizations can be considered also in the
description of the matter content of the shell, which describes the surface of the vacuum bubble.
In particular the presence of a surface tension is predicted by field theoretical approaches, which
describe the model in terms of scalar fields coupled to gravity (see the original ideas in
\cite{bib:PrThP1981..65..1443M,bib:PhReD1987..35..1747G,bib:PrThP1981..66..2052M,bib:PrThP1981..66..2287S,%
bib:PhLeB1982.108....98K,bib:PhLeB1982.108...103M,bib:PrThP1982..68..1979S} although there are, of course,
recent studies too \cite{bib:PhReD2006..74123520P,bib:PhReD2007..75103506P}):
also in this case, a preliminary analysis shows that the results in \cite{bib:gr-qc2007..06..1233G} survive
the addition of a surface tension, $\sigma$, to the model. A study of the consequences of the introduction of magnetic
monopoles for the conclusions of, both, \cite{bib:gr-qc2007.105....03G} and \cite{bib:gr-qc2007..06..1233G} is
also currently in progress \cite{bib:IPREP2007ShiGueAns..}, and seems to confirm that their properties are not
an accidental effects of a choice of a very particular model: thus they will likely survive under more general
conditions. Finally, it is clear that the above mentioned generalizations must address the problem beyond
the classical level, since quantum effects, besides being interesting in themselves, certainly play
a non negligible role in the universe creation/formation process. In view of the still unsolved
difficulties in the description of spacetime tunnelling \cite{bib:FFP6p2006........69S,%
bib:PhReD2005..72103525J,bib:PhReD2006..73123529J,bib:JGR162007...1...114A},
either in connection with child universes creation or not \cite{bib:QGF072007...1...004A},
it is certainly interesting to consider the feedback that the studies of these related aspects
might produce on each other. This would open the possibility to consider, on a firmer ground and
in a consistent framework, the suggestive phenomenological consequences of child universe
creation, particularly if unsuppressed. Some of this possibilities, which have been shortly
addressed in \cite{bib:gr-qc2007..06..1233G}, are under consideration and will also be
discussed in future work.

\section{\label{sec:conclu}Synopsis and discussion}

In this contribution, we have discussed child universe formation, reviewing in an historical perspective
the relevance and distinctive features of the process. We have also described a refinement
of one of the simplest realizations of a region of spacetime (child universe) disconnecting from an ambient one
(parent universe), namely the one in which the interior region is described by Minkowskii spacetime and the exterior
one by Schwarzschild spacetime: in our model, the consideration of a string gas for the matter composing the thin shells
that separates the two domains, supports the interesting idea that child universe creation can take place out of almost
empty space. We have also anticipated that natural generalizations of this model preserve this suggestive result. Summarizing,
the main picture that emerges from the previous discussion is as follows. Firstly, although we certainly
have in mind applications to the early universe cosmology, in this paper we have used the term \emph{child universe}
in a much more generic sense, as a region of spacetime that eventually disconnect from a pre-existing \emph{parent} one.
We have also seen that this process can take place starting from almost empty space. Moreover child universes which are
characterized by what we would like to call \emph{higher internal energy} seem to be produced at higher rate and could have
smaller action than those having a lower internal energy. In this sense it seems that production of what, ordinarily,
would be called excited baby universes is more likely than the production of ground state ones. Further
developments of these ideas and a more detailed account of their possible applications will be reported
elsewhere.

\section*{Acknowledgements}

The work of SA is supported by a long term invitation fellowship of
the Japan Society for the Promotion of Science (JSPS).


\begin{thebibliography}{99}

\bibitem{bib:PhReD2005..72103525J}
A.~{Aguirre} and M.~C. {Johnson}.
\newblock {{D}ynamics and Instability of False Vacuum Bubbles}.
\newblock {\em {Phys. Rev. D}}, 72:103525, 2005.

\bibitem{bib:PhReD2006..73123529J}
A.~{Aguirre} and M.~C. {Johnson}.
\newblock {{T}wo tunnels to inflation}.
\newblock {\em {Phys. Rev. D}}, 73:123529, 2006.

\bibitem{bib:PhReL1982..48..1220S}
A.~{Albrecht} and P.~J. {Steinhardt}.
\newblock {{C}osmology for Grand Unified Theories with Radiatively Induced
  Symmetry Breaking}.
\newblock {\em {Phys. Rev. Lett.}}, 48:1220, 1982.

\bibitem{bib:PhReD2000..62103509E}
S.~{Alexander}, R.~H. {Brandenberger}, and D.~{Easson}.
\newblock {{B}rane gases in the early universe}.
\newblock {\em {Phys. Rev. D}}, 62:103509, 2000.

\bibitem{bib:ClQuG2002..19..6321A}
S.~{Ansoldi}.
\newblock {{W}KB metastable quantum states of a de Sitter-Rei\ss{}ner-Nordstr\o{}m
  dust shell}.
\newblock {\em {Class. Quantum Grav.}}, 19:6321, 2002.

\bibitem{bib:AIPcp2005.751...159A}
S.~{Ansoldi}.
\newblock {{M}inisuperspace, WKB, quantum states of general relativistic
  extended objects}.
\newblock {\em {AIP Conf. Proc.}}, 751:159, 2005.

\bibitem{bib:JGR162007...1...114A}
S.~{Ansoldi}.
\newblock {{B}ubbles and quantum tunnelling in inflationary cosmology}.
\newblock In {\em 16th Workshop on General Relativity and Gravitation (JGRG16),
  Niigata, Japan, 27 Nov - 1 Dec 2006}, 2007.

\bibitem{bib:QGF072007...1...004A}
S.~{Ansoldi}.
\newblock {{V}acuum and semiclassical gravity: a difficulty and its bewildering
  significance}.
\newblock In {\em From Quantum to Emergent Gravity: Theory and Phenomenology,
  June 11-15 2007, Trieste, Italy}, 2007.

\bibitem{bib:ClQuG1997..14..2727S}
S.~{Ansoldi}, A.~{Aurilia}, R.~{Balbinot}, and E.~{Spallucci}.
\newblock {{C}lassical and Quantum Shell Dynamics and Vacuum Decay}.
\newblock {\em {Class. Quantum Grav.}}, 14:2727, 1997.

\bibitem{bib:gr-qc2007..06..1233G}
S.~{Ansoldi} and E.~I. {Guendelman}.
\newblock {{U}niverses out of almost empty space}.
\newblock arXiv:0706.1233 [gr-qc], 2007.

\bibitem{bib:IPREP2007ShiGueAns..}
S.~{Ansoldi}, E.~I. {Guendelman}, and I.~{Shilon}.
\newblock {{U}nsuppressed child universe production models}.
\newblock In preparation, 2007.

\bibitem{bib:FFP6p2006........69S}
S.~{Ansoldi} and L.~{Sindoni}.
\newblock {{G}ravitational tunnelling of relativistic shells}.
\newblock In {\em Proceedings of the 6th Internatinal Symposium of Frontiers in
  Fundamantal Physics, Udine, Italy, 26-29 Sep 2004}, 2006.

\bibitem{bib:PhReD1989..40..2511S}
A.~{Aurilia}, M.~{Palmer}, and E.~{Spallucci}.
\newblock {{E}volution of Bubbles in a Vacuum}.
\newblock {\em {Phys. Rev. D}}, 40:2511, 1989.

\bibitem{bib:PhReD1996..53..3215F}
C.~{Barrabes} and V.~P. {Frolov}.
\newblock {{H}ow many new worlds are inside a black hole?}
\newblock {\em {Phys. Rev. D}}, 53:3215, 1996.

\bibitem{bib:PhReD1991..43..1129I}
C.~{Barrabes} and W.~{Israel}.
\newblock {{T}hin Shells in General Relativity and Cosmology: the Lightlike
  Limit}.
\newblock {\em {Phys. Rev. D}}, 43:1129, 1991.

\bibitem{bib:PhReD2005..71064008I}
C.~{Barrabes} and W.~{Israel}.
\newblock {{L}agrangian brane dynamics in general relativity and in
  Einstein-Gauss-Bonnet gravity}.
\newblock {\em {Phys. Rev. D}}, 71:064008, 2005.

\bibitem{bib:PhReL1989..63...341V}
M.~{Barriola} and A.~{Vilenkin}.
\newblock {{G}ravitational field of a global monopole}.
\newblock {\em {Phys. Rev. Lett.}}, 63:341, 1989.

\bibitem{bib:PhReD1987..36..2919T}
V.~A. {Berezin}, V.~A. {Kuzmin}, and I.~I. {Tkachev}.
\newblock {{D}ynamics of Bubbles in General Relativity}.
\newblock {\em {Phys. Rev. D}}, 36:2919, 1987.

\bibitem{bib:PhReD1991..43.R3112T}
V.~A. {Berezin}, V.~A. {Kuzmin}, and I.~I. {Tkachev}.
\newblock {{B}lack holes initiate false-vacuum decay}.
\newblock {\em {Phys. Rev. D}}, 43:R3112--R3116, 1991.

\bibitem{bib:PhReD1987..35..1747G}
S.~K. {Blau}, E.~I. {Guendelman}, and A.~H. {Guth}.
\newblock {{D}ynamics of False Vacuum Bubbles}.
\newblock {\em {Phys. Rev. D}}, 35:1747, 1987.

\bibitem{bib:PhReD1999..59043513V}
A.~{Borde}, M.~{Trodden}, and T.~{Vachaspati}.
\newblock {{C}reation and structure of baby universes in monopole collisions}.
\newblock {\em {Phys. Rev. D}}, 59:043513, 1999.

\bibitem{bib:NuPhy2002B623...421K}
R.~{Brandenberger}, D.~A. {Easson}, and D.~{Kimberly}.
\newblock {{L}oitering phase in brane gas cosmology}.
\newblock {\em {Nucl. Phys.}}, B623:421, 2002.

\bibitem{bib:NuPhy1989B316...391V}
R.~H. {Brandenberger} and C.~{Vafa}.
\newblock {{S}uperstrings in the Early Universe}.
\newblock {\em {Nucl. Phys.}}, B316:391, 1989.

\bibitem{bib:PhReD1977..16..1762C}
J.~C.~G. {Callan} and S.~{Coleman}.
\newblock {{F}ate of the False Vacuum. II. First Quantum Corrections}.
\newblock {\em {Phys. Rev. D}}, 16:1762, 1977.

\bibitem{bib:PhReD1997..56..7621V}
I.~{Cho} and A.~{Vilenkin}.
\newblock {{S}pacetime structure of an inflating global monopole}.
\newblock {\em {Phys. Rev. D}}, 56:7621, 1997.

\bibitem{bib:PhReD1977..15..2929C}
S.~{Coleman}.
\newblock {{F}ate of the False Vacuum: Semiclassical Theory}.
\newblock {\em {Phys. Rev. D}}, 15:2929, 1977.

\bibitem{bib:PhReD1977..16..1248C}
S.~{Coleman}.
\newblock {{F}ate of the False Vacuum: Semiclassical Theory (errata)}.
\newblock {\em {Phys. Rev. D}}, 16:1248, 1977.

\bibitem{bib:PhReD1980..21..3305L}
S.~{Coleman} and F.~D. {Luccia}.
\newblock {{G}ravitational Effects On and Of Vacuum Decay}.
\newblock {\em {Phys. Rev. D}}, 21:3305, 1980.

\bibitem{bib:PhLeB2001.506...351G}
I.~G. {Dymnikova}, A.~{Dobosz}, M.~L. {Filchenkov}, and A.~{Gromov}.
\newblock {{U}niverses inside a Lambda black hole}.
\newblock {\em {Phys. Lett. B}}, 506:351, 2001.

\bibitem{bib:PhLeB1987.183...149G}
E.~{Farhi} and A.~A. {Guth}.
\newblock {{A}n obstacle to creating a universe in the laboratory}.
\newblock {\em {Phys. Lett. B}}, 183:149, 1987.

\bibitem{bib:NuPhy1990B339...417G}
E.~{Farhi}, A.~H. {Guth}, and J.~{Guven}.
\newblock {{I}s it possible to create a universe in the laboratory by quantum
  tunneling?}
\newblock {\em {Nucl. Phys.}}, B339:417, 1990.

\bibitem{bib:PhReD1990..42..4042P}
W.~{Fischler}, D.~{Morgan,}, and J.~{Polchinski}.
\newblock {{Q}uantization of false-vacuum bubbles: A Hamiltonian treatment of
  gravitational tunneling}.
\newblock {\em {Phys. Rev. D}}, 42:4042, 1990.

\bibitem{bib:PhReD1990..41..2638P}
W.~{Fischler}, D.~{Morgan}, and J.~{Polchinski}.
\newblock {{Q}uantum Mechanics of False Vacuum Bubbles}.
\newblock {\em {Phys. Rev. D}}, 41:2638, 1990.

\bibitem{bib:gr-qc2007.105....03G}
E.~I. {Guendelman}.
\newblock {{C}hild universes UV regularization?}
\newblock gr-qc/0703105, 2007.

\bibitem{bib:ClQuG1999..16..3315P}
E.~I. {Guendelman} and J.~{Portnoy}.
\newblock {{T}he universe out of an elementary particle?}
\newblock {\em {Class. Quantum Grav.}}, 16:3315, 1999.

\bibitem{bib:MPhLA2001..16..1079P}
E.~I. {Guendelman} and J.~{Portnoy}.
\newblock {{A}lmost classical creation of a universe}.
\newblock {\em {Mod. Phys. Lett. A}}, 16:1079, 2001.

\bibitem{bib:PhReD1991..44..3152R}
E.~I. {Guendelman} and A.~{Rabinowitz}.
\newblock {{G}ravitational field of a hedgehog and the evolution of vacuum
  bubbles}.
\newblock {\em {Phys. Rev. D}}, 44:3152, 1991.

\bibitem{bib:PhReD1981..23...347G}
A.~H. {Guth}.
\newblock {{I}nflationary universe: A possible solution to the horizon and
  flatness problems}.
\newblock {\em {Phys. Rev. D}}, 23:347, 1981.

\bibitem{bib:PhReL1965..15...689H}
S.~W. {Hawking}.
\newblock {{O}ccurrence of singularities in open universes}.
\newblock {\em {Phys. Rev. Lett.}}, 15:689, 1965.

\bibitem{bib:PRSLA1966.294...511H}
S.~W. {Hawking}.
\newblock {{T}he occurrence of singularities in cosmology}.
\newblock {\em {Proc. Roy. Soc. Lon. A}}, 294:511, 1966.

\bibitem{bib:PRSLA1966.295...490H}
S.~W. {Hawking}.
\newblock {{T}he occurrence of singularities in cosmology. II}.
\newblock {\em {Proc. Roy. Soc. Lon. A}}, 295:490, 1966.

\bibitem{bib:PRSLA1967.300...187H}
S.~W. {Hawking}.
\newblock {{T}he occurrence of singularities in cosmology. III. Causality and
  singularities}.
\newblock {\em {Proc. Roy. Soc. Lon. A}}, 300:187, 1967.

\bibitem{bib:CaUPr1973...1...391E}
S.~W. {Hawking} and G.~F.~R. {Ellis}.
\newblock {\em {{T}he Large Scale Structure of Space-Time}}.
\newblock {Cambridge University Press}, 1973.

\bibitem{bib:PRSLA1970.314...529P}
S.~W. {Hawking} and R.~{Penrose}.
\newblock {{T}he singularities of gravitational collapse and cosmology}.
\newblock {\em {Proc. Roy. Soc. Lon. A}}, 314:529, 1970.

\bibitem{bib:NuCim1966.B44.....1I}
W.~{Israel}.
\newblock {{S}ingular Hypersurfaces and Thin Shells in General Relativity}.
\newblock {\em {Nuovo Cimento}}, B44:1, 1966.

\bibitem{bib:NuCim1967.B48...463I}
W.~{Israel}.
\newblock {{S}ingular Hypersurfaces and Thin Shells in General Relativity
  (Errata)}.
\newblock {\em {Nuovo Cimento}}, B48:463, 1967.

\bibitem{bib:PrThP1981..66..2052M}
H.~{Kodama}, M. {Sasaki}, K.~{Sato}, and K.-I. {Maeda}.
\newblock {{F}ate of Wormholes Created by 1st Order Phase-Transiitons in the
  Early Universe}.
\newblock {\em {Progr. Theor. Phys.}}, 66:2052, 1981.

\bibitem{bib:PrThP1982..68..1979S}
H.~{Kodama}, M.~{Sasaki}, and K.~{Sato}.
\newblock {{A}bundance of Primordial Holes Produced by Cosmological 1st-order
  Phase Transition}.
\newblock {\em {Progr. Theor. Phys.}}, 68:1979, 1982.

\bibitem{bib:AstrJ1989.344...543K}
E.~W. {Kolb}.
\newblock {{A} Coasting Cosmology}.
\newblock {\em {Astroph. J.}}, 344:543, 1989.

\bibitem{bib:PhReD2007..75103506P}
B.~H. {Lee}, W.~{Lee}, S.~{Nam}, and C.~{Park}.
\newblock {{D}omain wall cosmology and multiple accelerations}.
\newblock {\em {Phys. Rev. D}}, 75:103506, 2007.

\bibitem{bib:PhReD1987..36..1088W}
K.~{Lee} and E.~J. {Weinberg}.
\newblock {{D}ecay of True Vacuum in Curved SpaceTime}.
\newblock {\em {Phys. Rev. D}}, 36:1088, 1987.

\bibitem{bib:PhReD2006..74123520P}
W.~{Lee}, B.~H. {Lee}, C.~H. {Lee}, and C.~{Park}.
\newblock {{F}alse vacuum bubble nucleation due to a nonminimally coupled
  scalar field}.
\newblock {\em {Phys. Rev. D}}, 74:123520, 2006.

\bibitem{bib:PhLeB1994.327...208L}
A.~{Linde}.
\newblock {{M}onopoles as big as a universe}.
\newblock {\em {Phys. Lett. B}}, 327:208, 1994.

\bibitem{bib:PhLeB1982.108...389L}
A.~D. {Linde}.
\newblock {{A} new inflationary universe scenario: A possible solution of the
  horizon, flatness, homogeneity, isotropy and primordial monopole problems}.
\newblock {\em {Phys. Lett. B}}, 108:389, 1982.

\bibitem{bib:MPhLA1986...1....81L}
A.~D. {Linde}.
\newblock {{E}ternal Chaotic Inflation}.
\newblock {\em {Mod. Phys. Lett. A}}, 1:81, 1986.

\bibitem{bib:PhLeB1986.175...395L}
A.~D. {Linde}.
\newblock {{E}thernally existing self-reproducing chaotic inflationary
  universe}.
\newblock {\em {Phys. Lett. B}}, 175:395, 1986.

\bibitem{bib:PhLeB1982.108....98K}
K.-I. {Maeda}, K.~{Sato}, M.~{Sasaki}, and H.~{Kodama}.
\newblock {{C}reation of Sch\-warzschild-de Sitter wormholes by a cosmological
  first-order phase transition}.
\newblock {\em {Phys. Lett. B}}, 108:98, 1982.

\bibitem{bib:PhLeB1983.121...313P}
S.~{Parke}.
\newblock {{G}ravity and the decay of false vacuum}.
\newblock {\em {Phys. Lett. B}}, 121:313, 1983.

\bibitem{bib:PhReD1996..54..1548S}
N.~{Sakai}.
\newblock {{D}ynamics of gravitating magnetic monopoles}.
\newblock {\em {Phys. Rev. D}}, 54:1548, 1996.

\bibitem{bib:PhReD2006..74024026K}
N.~{Sakai}, K.-I. {Nakao}, H.~{Ishihara}, and M.~{Kobayashi}.
\newblock {{I}s it possible to create a universe out of a monopole in the
  laboratory?}
\newblock {\em {Phys. Rev. D}}, 74:024026, 2006.

\bibitem{bib:PhReD1996..53...655M}
N.~{Sakai}, H.-A. {Shinkai}, T.~{Tachizawa}, and K.-I. {Maeda}.
\newblock {{D}ynamics of topological defects and inflation}.
\newblock {\em {Phys. Rev. D}}, 53:655, 1996.

\bibitem{bib:PrThP1981..66..2287S}
K.~{Sato}.
\newblock {{P}roduction of Magnetized Black-Hole and Wormholes by 1st-order
  Phase Transitions in the Early Universe}.
\newblock {\em {Progr. Theor. Phys.}}, 66:2287, 1981.

\bibitem{bib:PhLeB1982.108...103M}
K.~{Sato}, H.~{Kodama}, M.~{Sasaki}, and K.-I. {Maeda}.
\newblock {{M}ulti-production of universes by first-order phase transition of a
  vacuum}.
\newblock {\em {Phys. Lett. B}}, 108:103, 1982.

\bibitem{bib:PrThP1981..65..1443M}
K.~{Sato}, M.~{Sasaki}, H.~{Kodama}, and K.-I. {Maeda}.
\newblock {{C}reation of Wormholes by 1st Order Phase-Transiton of a Vacuum in
  the Early Universe}.
\newblock {\em {Progr. Theor. Phys.}}, 65:1443, 1981.

\bibitem{bib:MPhLA1991...6..1535T}
A.~{Tomimatsu}.
\newblock {{C}ollapse of wormhole space and the baby universe production}.
\newblock {\em {Mod. Phys. Lett. A}}, 6:1535, 1991.

\bibitem{bib:PhReD1984..30...509V}
A.~{Vilenkin}.
\newblock {{Q}uantum Creation of Universes}.
\newblock {\em {Phys. Rev. D}}, 30:509, 1984.

\bibitem{bib:PhReL1994..72..3137V}
A.~{Vilenkin}.
\newblock {{T}opological Inflation}.
\newblock {\em {Phys. Rev. Lett.}}, 72:3137, 1994.

\bibitem{bib:UofCP1984...1...491W}
R.~M. {Wald}.
\newblock {\em {{G}eneral Relativity}}.
\newblock {The University of Chicago Press}, 1984.
\end{thebibliography}
\end{document}